\documentclass[11pt,letterpaper]{article}

\usepackage{theorem,latexsym,graphicx}
\usepackage{amsmath,amssymb,enumerate}
\usepackage{xspace}
\usepackage{fullpage}
\usepackage{url}
\usepackage{algorithm}
\usepackage[noend]{algorithmic}
\usepackage{subfigure}
\usepackage{verbatim}
\usepackage{times}
\usepackage{paralist}
\usepackage{multicol}
\usepackage{mathtools}

\newtheorem{theorem}{Theorem}[section]
\newtheorem{definition}[theorem]{Definition}
\newtheorem{lemma}[theorem]{Lemma}
\newtheorem{observation}[theorem]{Observation}
\newtheorem{corollary}[theorem]{Corollary}

\newenvironment{proof}{

\noindent{\bf Proof:}}
{\hfill$\blacksquare$

}

\newenvironment{proofof}[1]{

\noindent{\bf Proof of {#1}:}}
{\hfill$\blacksquare$

}

\newcommand{\junk}[1]{}
\newcommand{\ignore}[1]{}

\newcommand{\R}[0]{{\ensuremath{\mathbb{R}}}}

\newcommand{\OPT}{\ensuremath{\textrm{OPT}}}

\newcommand{\E}{\mathbb{E}}

\newcommand{\qedsymb}{\hfill{\rule{2mm}{2mm}}}

\newcommand{\initOneLiners}{%
    \setlength{\itemsep}{0pt}
    \setlength{\parsep }{0pt}
    \setlength{\topsep }{0pt}
}

\newcommand{\dpa}{\textsc{DPA}\xspace}
\newcommand{\dpai}{DPA\xspace}
\newcommand{\otp}{\textsc{OTP}\xspace}

\DeclareMathOperator{\skewm}{skewm}
\DeclareMathOperator{\skewp}{skewp}

\newcommand{\boldl}{v}

\newcommand{\myparagraph}[1]{\smallskip \noindent \emph{#1}}
\newcommand{\beforesection}{}
\newcommand{\aftersection}{}

\newcommand{\beforesubsection}{}
\newcommand{\balemma}{}

\begin{document}

\title{Geometry of Online Packing Linear Programs}
\author{Marco Molinaro \\ Carnegie Mellon \and R. Ravi \\Carnegie Mellon}
\date{}

\maketitle

\begin{abstract}
We consider packing LP's with $m$ rows where all constraint coefficients are normalized to be in the unit interval. The $n$ columns arrive in random order and the goal is to set the corresponding decision variables irrevocably when they arrive so as to obtain a feasible solution maximizing the expected reward. Previous $(1 - \epsilon)$-competitive algorithms require the right-hand side of the LP to be $\Omega (\frac{m}{\epsilon^2} \log \frac{n}{\epsilon})$, a bound that worsens with the number of columns and rows. However, the dependence on the number of columns is not required in the single-row case and known lower bounds for the general case are also independent of $n$. 

Our goal is to understand whether the dependence on $n$ is required in the multi-row case, making it fundamentally harder than the single-row version. We refute this by exhibiting an algorithm which is $(1 - \epsilon)$-competitive as long as the right-hand sides are $\Omega (\frac{m^2}{\epsilon^2} \log \frac{m}{\epsilon})$. Our techniques refine previous PAC-learning based approaches which interpret the online decisions as linear classifications of the columns based on sampled dual prices. The key ingredient of our improvement comes from a non-standard covering argument together with the realization that only when the columns of the LP belong to few 1-d subspaces we can obtain small such covers; bounding the size of the cover constructed also relies on the geometry of linear classifiers. General packing LP's are handled by perturbing the input columns, which can be seen as making the learning problem more robust.  
\end{abstract}


	\beforesection
	\section{Introduction}
	\aftersection
	
Traditional optimization models usually assume that the input is known a priori. However, in most applications, the data is either revealed over time or only coarse information about the input is known, often modeled in terms of a probability distribution. Consequently, much effort has been directed towards understanding the quality of solutions that can be obtained without full knowledge of the input, which led to the development of online and stochastic optimization \cite{borodinbook,stocprogbook}. Emerging problems such as allocating advertisement slots to advertisers and yield management in the internet are of inherent online nature and have further accelerated this development \cite{agrawal}.
	
Linear programming is arguably the most important and thus well-studied optimization problem. Therefore, understanding the limitations of solving linear programs when complete data is not available is a fundamental theoretical problem with a slew of applications, including the ad allocation and yield management problems above. Indeed, a simple linear program with one uniform knapsack constraint, the Secretary Problem, was one of the first online problems to be considered and an optimal solution was already obtained by the early 60's \cite{Dynkin,GilbertMosteller}. Although the single knapsack case is currently well-understood under different models of how information is revealed \cite{BabaioffSurvey}, much less is known about problems with multiple knapsacks and only recently algorithms with solution guarantees have been developed \cite{feldman,agrawal,Devanur11}.

\myparagraph{The Model.} We study online packing LP's in the \emph{random permutation model}. Consider a fixed but unknown LP with $n$ columns $a^1, a^2, \ldots, a^n \in [0,1]^m$, whose associated variables are constrained to be in $[0,1]$, and $m$ packing constraints:
	%
	\begin{align}
		\OPT = \max \sum_{t = 1}^n \pi_t x_t \notag \\
		\sum_{t = 1}^n a^t x_t \le B \label{eq:LP} \tag{LP}\\
		x_t \in [0,1] \,. \notag
	\end{align}
	Columns are presented in uniformly random order, and when a column is presented we are required to irrevocably choose the value of its corresponding variable. We assume that the number of columns $n$ is known.\footnote{Actually knowing $n$ up to $(1 \pm \epsilon)$ factor is enough. This assumption is required to allow algorithms with non-trivial competitive ratio \cite{DevanurHayes09}.} The goal is to obtain a feasible solution while maximizing its value. We use $\OPT$ to denote the optimum value of the (offline) LP.
	
	By scaling down rows as necessary, we assume without loss of generality that all entries of $B$ are the same, which we also denote with some overload of notation by $B$. Due to the packing nature of the problem, we also assume without loss of generality that all the $\pi_t$'s are non-negative and all the $a^t$'s are non-zero: we can simply ignore columns which do not satisfy the first property and always set to 1 the variables associated to the remaining columns which do not satisfy the second property. Finally, we assume that the columns $a^t$'s are in \emph{general position}: for all $p \in \mathbb{R}^m$, there are at most $m$ different $t \in [n]$ such that $\pi_t = p a^t$. Notice that perturbing the input randomly by a tiny amount achieves this property with probability one, while the effect of the perturbation is absorbed in our approximation guarantees \cite{DevanurHayes09,agrawal}.

\myparagraph{Related work.} The random permutation model has grown in popularity~\cite{GoelMehta08,DevanurHayes09,BabaioffSurvey} since it avoids strong lower bounds of the pessimistic adversarial-order model~\cite{BuchbinderMOR} while still capturing the lack of total information a priori. Different online problems have already been studied in this model, including bin-packing \cite{kenyon}, matchings \cite{KVV,GoelMehta08}, the AdWords Problem \cite{DevanurHayes09} and different generalizations of the Secretary Problem \cite{BabaioffSurvey,weightsSecretary,submodularSecretary,soto,sungjin}. Closest to our work are packing problems with a single knapsack constraint. In \cite{kleinberg}, Kleinberg considered the $B$-Choice Secretary Problem, where the goal is to select at most $B$ items coming online in random order to maximize profit. The author presented an algorithm with competitive ratio $1 - O(1/\sqrt{B})$ and showed that $1 - \Omega(1/\sqrt{B})$ is best possible. Generalizing the $B$-Choice Secretary Problem, Babaioff et al. \cite{babaioff} considered the online knapsack problem and presented a $(1/10e)$-competitive algorithm. Notice that in both cases the competitive ratio does not depend on $n$.

Despite all these works, the first result for more general online packing LP's here was only recently obtained by Feldman et al.~\cite{feldman} and Agrawal et al. \cite{agrawal}. The first paper presents an algorithm that obtains with high probability a solution of value at least $(1 - \epsilon) \OPT$ whenever $B \ge \Omega(\frac{m \log n}{\epsilon^3})$ and $\OPT \ge \Omega(\frac{\pi_{\max} m \log n}{\epsilon})$, where $\pi_{\max}$ is the largest profit. In the second paper, the authors present an algorithm which obtains a solution of expected value at least $(1 - \epsilon) \OPT$ under the weaker assumptions $B \ge \Omega\left(\frac{m}{\epsilon^2} \log \frac{n}{\epsilon}\right)$ or $\OPT \ge \Omega\left(\frac{\pi_{\max} m^2}{\epsilon^2} \log \frac{n}{\epsilon}\right)$. One other way of stating this result is that the algorithm obtains a solution with competitive ratio $1 - O(\sqrt{\frac{m \log(n) \log B}{B}})$; notice that the guarantee degrades as $n$ increases. The current lower bound on $B$ to allow $(1 - \epsilon)$-competitive algorithms is $B \ge \frac{\log m}{\epsilon^2}$, also presented in \cite{agrawal}. We remark that these algorithms actually work for more general allocation problems, where a set of columns representing various options arrive at each step and the solution may choose at most one of the options. 
	
Both of the above algorithms use a connection between solving the online LP and PAC-learning \cite{cucker} a linear classification of its columns, which was initiated by Devanur and Hayes~\cite{DevanurHayes09} in the context of the AdWords problem. Here we further explore this connection and our improved bounds can be seen as a consequence of making the learning algorithm more robust by suitably changing the input LP. Robustness is a topic well-studied in learning theory \cite{devroye,partha}, although existing results do not seem to apply directly to our problem. We remark that a component of robustness more closely related to the standard PAC-learning literature is used in \cite{DevanurHayes09}.
	
In recent work, Devanur et al.~\cite{Devanur11} consider the weaker \emph{i.i.d. model} for the general allocation problem. While in the random permutation model one can think that columns are sampled without replacement, in the i.i.d. model they are sampled with replacement. Making use of the independence between samples, Devanur et al. substantially improve requirement on $B$ to $\Omega(\frac{\log (m/\epsilon)}{\epsilon^2})$ while showing that the lower bound $\Omega\left(\frac{\log m}{\epsilon^2}\right)$ still holds in this model. We remark, however, that these models can present very different behaviors: as a simple example, consider an LP with $n$ columns, $m = 1$ constraints and budget $B = 1$, where only one of the columns has $\pi_1 = a^1 = 1$ and all others have $\pi_i = a^i = 0$; in the random permutation model the expected value of the optimal solution is 1, while in the i.i.d. model this value is $1 - (1 - 1/n)^n \rightarrow 1 - 1/e$. The competitiveness of the algorithm of \cite{Devanur11} under the permutation model is still unknown and was left as an open problem by the authors. 
	
\myparagraph{Our results.} Our focus is to understand how large $B$ is required to be in order to allow $(1 - \epsilon)$-competitive algorithms. In particular, the requirements for $B$ in the above algorithms degrade as the number of columns in the LP increases, while the the lower bound does not. With the trend of handling LP's with larger number of columns (e.g. columns correspond to the keywords in the ad allocation problem, which in turn correspond to visits of a search engine's webpage),
this gap is very unsatisfactory from a practical point of view. Furthermore, given that guarantees for the single knapsack case do not depend on the number of columns, it is important to understand if the multi-knapsack case is fundamentally more difficult. In this work, we give a precise indication of why the latter problem was resistant to arguments used in the single knapsack case, and overcome this difficulty to exhibit an algorithm with dimension-independent guarantee.

We show that a modification of the \dpa algorithm from~\cite{agrawal} that we call \emph{Robust \dpai} obtains a $(1 -\epsilon)$-competitive solution for online packing LP's with $m$ constraints in the random permutation model whenever $B \ge \Omega(\frac{m^2}{\epsilon^2} \log \frac{m}{\epsilon})$. Another way of stating this result is that the algorithm has competitive ratio $1 - O(m \sqrt{\log B}/\sqrt{B})$. Contrasting to previous results, our guarantee does not depend on $n$ and in the case $m = 1$ matches the bounds for the $B$-Choice Secretary Problem up to lower order terms. We finally remark that we can replace the requirement $B \ge \Omega(\frac{m^2}{\epsilon^2} \log \frac{m}{\epsilon})$ by $\OPT \ge \Omega(\frac{\pi_{\max} m^3}{\epsilon^2} \log \frac{m}{\epsilon})$ exactly as done in Section 5.1 of \cite{agrawal}. 


\myparagraph{High-level outline.} As mentioned before, we use the connection between solving an online LP and PAC-learning a good linear classification of its columns; in order to obtain the improved guarantee, we focus on tightening the bounds for the generalization error of the learning problem. More precisely, solving the LP can be seen as classifying the columns into 0/1, which corresponds to setting their associated variable to 0/1. Consider a family $\mathcal{X} \subseteq \{0,1\}^n$ of linear classifications of the columns. Our algorithms sample a set $S$ of columns and learn a classification $x^S \in \mathcal{X}$ which is ``good'' for the columns $S$ (i.e., obtains large proportional revenue while not filling up the proportionally scaled budget too much). The goal is to upper bound the probability that $x^S$ is not good for the whole LP; this is typically done via a union bound over the classifications in $\mathcal{X}$ \cite{DevanurHayes09,agrawal}.

To obtain improved guarantees, we refine this bound using an argument akin to covering: we consider \emph{witnesses} (Section \ref{sec:witness}), which are representatives of groups of `similar' bad classifications that can be used to bound the probability that \emph{any} classification in the group is learned; for that we need to use a non-standard measure of similarity between classifications which is based on the budget of the LP. The problem is that, when the columns $(\pi_t, a^t)$'s do not lie in a two-dimensional subspace of $\mathbb{R}^{m}$, the set $\mathcal{X}$ may contain a large number of mutually dissimilar bad classifications; this is a roadblock for obtaining a small set of witnesses. In stark contrast, when these columns do lie in a two-dimensional subspace (e.g., $m = 1$), these classifications have a much nicer structure which indeed allows a small set of witnesses. This indicates that the latter learning problem is intrinsically more robust than the former, which seem to precisely capture the increased difficulty in obtained good bounds for the multi-row case.

Motivated by this discussion we first consider LP's whose columns $a^t$'s lie in \emph{few} one-dimensional subspaces (Section~\ref{sec:otp}). For each of these subspaces, we are able to approximate the classifications induced in the columns lying in the subspace by considering a small subset of the induced classifications; patching together these partial classifications gives us a witness set for $\mathcal{X}$. However, this strategy as stated does not make use of the fact that the subspaces are embedded in an $m$-dimensional space, and hence leads to large witness sets. By establishing a connection between the ``useful'' patching possibilities with faces of a hyperplane arrangement in $\mathbb{R}^m$ (Lemma \ref{lemma:sizeP}), we are able to make use of the dimension of the host space and exhibit witness sets of much smaller sizes, which leads to improved bounds.

For a general packing LP, we perturb the columns $a^t$'s to make them lie in few one-dimensional subspaces that form an `$\epsilon$-net' of the space, while not altering the feasibility and optimality of the LP by more than a $(1 \pm \epsilon)$ factor (Section~\ref{sec:rotp}). Finally, we tighten the bound by using the idea of periodically recomputing the classification, following~\cite{agrawal} (Section~\ref{sec:rdpa}).

\beforesection
\section{OTP for almost 1-dim columns} \label{sec:otp}
\aftersection

	In this section we describe and analyze the algorithm \otp (One-Time Pricing) over LP's whose columns are contained in few 1-dimensional subspaces of $\mathbb{R}^{m}$. The overall goal is to find an appropriate dual (perhaps infeasible) solution $p$ for \eqref{eq:LP} and use it to classify the columns of the LP. More precisely, given $p \in \mathbb{R}^m$, we define $x(p)_t = 1$ if $\pi_t > p a^t$ and $x(p)_t = 0$ otherwise. Thus, $x(p)$ is the result of classifying the columns $(\pi_t, a^t)$'s with the homogeneous hyperplane in $\mathbb{R}^{m+1}$ with normal $(-1, p)$. The motivation behind this classification is that it selects the columns which have positive reduced cost with respect to the dual solution $p$, or alternatively, it solves to optimality the Lagrangian relaxation using $p$ as multipliers.
	
 \myparagraph{Sampling LP's.} In order to obtain a good dual solution $p$ we use the (random) LP consisting on the first $s$ columns of \eqref{eq:LP} with appropriately scaled right-hand side.
	%
%
	%
	
	\hspace{-1.5cm}
	\begin{minipage}[b]{0.5\linewidth}
		\begin{align} \tag{$(s, \delta)$-LP} \label{eq:sdLP}
			\max & \sum_{t = 1}^{s} \pi_{\sigma(t)} x_{\sigma(t)} \\
			&\sum_{t = 1}^{s} a^{\sigma(t)} x_{\sigma(t)} \le \frac{s}{n} \delta B \notag \\
			&x_{\sigma(t)} \in [0,1] \ \ \ \ t = 1, \ldots, s \notag .
		\end{align}
	\end{minipage}
	\hspace{0.25cm}
	\vrule
	\hspace{0.25cm}
	\begin{minipage}[b]{0.5\linewidth}
		\begin{align} \tag{$(s, \delta)$-Dual} \label{eq:sdDual}
			\min \ & \frac{s}{n} \delta B \sum_{i = 1}^{m} p_i + \sum_{t = 1}^s \alpha_{\sigma(t)} \\
			& p a^{\sigma(t)} + \alpha_{\sigma(t)} \ge \pi_{\sigma(t)} \ \ \ \ t = 1, \ldots, s \notag  \\
			& p \ge 0 \notag \\
			& \alpha \ge 0 \notag.
		\end{align}
	\end{minipage}
\ignore{	
  \setlength{\columnsep}{18pt}
	\begin{multicols}{2}
		\setlength{\columnseprule}{.4pt}
		\begin{align} \tag{$(s, \delta)$-LP} \label{eq:sdLP}
			\max & \sum_{t = 1}^{s} \pi_{\sigma(t)} x_{\sigma(t)} \\
			&\sum_{t = 1}^{s} a^{\sigma(t)} x_{\sigma(t)} \le \frac{s}{n} \delta B \notag \\
			&x_{\sigma(t)} \in [0,1] \ \ \ \ t = 1, \ldots, s \notag .
		\end{align}
		\begin{align} \tag{$(s, \delta)$-Dual} \label{eq:sdDual}
			\min \ & \frac{s}{n} \delta B \sum_{i = 1}^{m} p_i + \sum_{t = 1}^s \alpha_{\sigma(t)} \\
			& p a^{\sigma(t)} + \alpha_{\sigma(t)} \ge \pi_{\sigma(t)} \ \ \ \ t = 1, \ldots, s \notag  \\
			& p \ge 0 \notag \\
			& \alpha \ge 0 \notag.
		\end{align}
	\end{multicols}
}

\noindent Here $\sigma$ denotes the random permutation of the columns of the LP. We use $\OPT(s,\delta)$ to denote the optimal value of $(s, \delta)$-LP and $\OPT(s)$ to denote the optimal value of $(s, 1)$-LP.
	
	The static pricing algorithm \otp of \cite{agrawal} can then be described as follows.\footnote{To simplify the exposition, we assume that $\epsilon n$ is an integer.}

\begin{enumerate}
 \item Wait for the first $\epsilon n$ columns of the LP (indexed by $\sigma(1), \sigma(2), \ldots, \sigma(\epsilon n)$) and solve $(\epsilon n, 1 - \epsilon)$-Dual. Let $(p, \alpha)$ be the obtained dual optimal solution.
 \item Use the classification given by $p$ as above by setting $x_{\sigma(t)} = x(p)_{\sigma(t)}$ for $t = \epsilon n + 1, \epsilon n + 2, \ldots$ for as long as the solution obtained remains valid. From this point on set all further variables to zero.
\end{enumerate}

	Note that by definition this algorithm outputs a feasible solution with probability one. Our goal is then to analyze the quality of the solution produced, ultimately leading to the following theorem.
	
	\balemma
	\begin{theorem} \label{thm:otp}
		Fix $\epsilon \in (0,1]$. Suppose that there are $K \ge m$ 1-dim subspaces of $\mathbb{R}^m$ containing the columns $a^t$'s and that $B \ge \Omega\left(\frac{m}{\epsilon^3} \log \frac{K}{\epsilon}\right)$. Then algorithm \otp returns a feasible solution with expected value at least $(1 - 5\epsilon) \OPT$.
	\end{theorem}
	\balemma
	
	Let $S = \{\sigma(1), \ldots, \sigma(\epsilon n)\}$ be the (random) index set of the columns sampled by \otp. We use $p^S$ to denote the optimal dual solution obtained by \otp; notice that $p^S$ is completely determined by $S$. To simplify the notation, we also use $x^S$ to denote $x(p^S)$.
	
		Notice that, for all the scenarios where $x^S$ is feasible, the solution returned by \otp  is identical to $x^S$ with its components $x^S_{\sigma(1)}, \ldots, x^S_{\sigma(\epsilon n)}$ set to zero. Given this observation and the fact that $\E[\sum_{t \leq \epsilon n} \pi_{\sigma(t)} x^S_{\sigma(t)}] \leq \epsilon \OPT$, one can prove that the following lemma implies Theorem \ref{thm:otp}.
%
	
	\balemma
	\begin{lemma} \label{lemma:goodOtp}
		Fix $\epsilon \in (0,1]$. Suppose that there are $K \ge m$ 1-dim subspaces of $\mathbb{R}^m$ containing the columns $a^t$'s and that $B \ge \Omega\left(\frac{m}{\epsilon^3} \log \frac{K}{\epsilon}\right)$. Then with probability at least $(1 - \epsilon)$, $x^S$ is a feasible solution for \eqref{eq:LP} with value at least $(1 - 3\epsilon) \OPT$.
	\end{lemma}
	\balemma

\beforesubsection
	\subsection{Connection to PAC learning}
	\aftersection
	
	We assume from now on that $B \ge \Omega(\frac{m}{\epsilon^3} \log \frac{K}{\epsilon})$. Let $\mathcal{X} = \{x(p) : p \in \mathbb{R}^m_+\} \subseteq \{0,1\}^n$ denote the set of all possible linear classifications of the LP columns which can be generated by \otp. With slight overload in the notation, we identify a vector $x \in \{0,1\}^n$ with the subset of $[n]$ corresponding to its support.

\balemma
\begin{definition}[Bad solution]
Given a scenario, we say that $x^S$ is \emph{bad} if it does not satisfy the properties of Lemma \ref{lemma:goodOtp}, namely $x^S$ is either infeasible or has value less than $(1 - 3\epsilon) \OPT$. We say that $x^S$ is \emph{good} otherwise.
\end{definition}
\balemma
	
	As noted in previous work, since our decisions are made based on reduced costs it suffices to analyze the \emph{budget occupation} (or complementary slackness) of the solution in order to understand its \emph{value}. To make this precise, given $x \in \{0,1\}^n$ let $a_i(x) = \sum_{t \in x} a_i^t$ be its occupation of the $i$th budget and let $a^S_i(x) = \frac{1}{\epsilon}\sum_{t \in x \cap S} a_i^t$ be its appropriately scaled occupation of $i$th budget in the sampled LP (recall $|S| = \epsilon n$).
	
	\balemma
	\begin{lemma} \label{lemma:approximateCS}
	 	Consider a scenario where $x^S$ satisfies: (i) for all $i \in [m]$, $a_i(x^S) \le B$ and (ii) for all $i \in [m]$ with $p^S_i > 0$, $a_i(x^S) \ge (1 - 3 \epsilon) B$. Then $x^S$ is good.
	\end{lemma}
	\balemma
	
	Moreover, since we are making decisions based on the \emph{optimal} reduced cost for the sampled LP, our solution satisfies the above properties for the sampled LP. 
	
	
	\balemma
	\begin{lemma} \label{lemma:sampleCS}
		In every scenario, $x^S$ satisfies the following: (i) for all $i \in [m]$, $a_i^S(x^S) \le (1 - \epsilon)B$ and (ii) for every $i \in [m]$ with $p^S_i > 0$, $a_i^S(x^S) \ge (1 - 2\epsilon) B$.
	\end{lemma}
	\balemma
	


	Given that $a_i(x) = \E[a^S_i(x)]$ for all $x$, the idea is to use concentration inequalities to argue that the conditions in Lemma \ref{lemma:approximateCS} hold with good probability. Although concentration of $a^S_i(x)$ for \emph{fixed} $x$ can be achieved via Chernoff-type bounds, the quantity $a^S_i(x^S)$ has undesired correlations; obtaining an effective bound is the main technical contribution of this paper. 
	
\balemma
\begin{definition}[Badly learnable]
	For a given scenario, we say that $x \in \mathcal{X}$ can be \emph{badly learned for budget $i$} if either (i) $a_i^S(x) \le (1 - \epsilon) B$ and $a_i(x) > B$ or (ii) $a^S_i(x) \ge (1 - 2\epsilon) B$ and $a_i(x) < (1 - 3 \epsilon) B$.
\end{definition}
\balemma

	 Essentially these are the classifications which look good for the sampled $(\epsilon n, 1- \epsilon)$-LP but are actually bad for \eqref{eq:LP}. Putting Lemmas~\ref{lemma:approximateCS} and \ref{lemma:sampleCS} together and unraveling the definitions gives that 
 	%
	\begin{equation*}
		\Pr\left(x^S \textrm{ is bad}\right) \le \Pr\left(\bigvee_{i \in [m], x \in \mathcal{X}} x \textrm{ can be badly learned for budget } i\right). \label{eq:badlyLearned}
	\end{equation*}
		Notice that the right-hand side of this inequality does not depend on $x^S$, it is only a function of how skewed $a_i^S(x)$ is as compared to its expectation $a_i(x)$.
	
	Usually the right-hand side in the previous equation is upper bounded by taking a union bound over all its terms \cite{agrawal}. Unfortunately this is too wasteful: when $x$ and $x'$ are ``similar'' there is a large overlap between the scenarios where $a_i^S(x)$ is skewed and those where $a_i^S(x')$ is skewed. In order to obtain improved guarantees, we introduce in the next section a new way of bounding the right-hand side of the above expression. 
%
	
	\beforesubsection
	\subsection{Similarity via witnesses} \label{sec:witness}
	\aftersection
	
	First, we partition the classifications which can be badly learned for budget $i$ into two sets, depending on why they are bad: for $i \in [m]$, let $\mathcal{X}_i^+ = \{x \in \mathcal{X} : a_i(x) > B\}$ and $\mathcal{X}_i^- = \{x \in \mathcal{X} : a_i(x) < (1 - 3 \epsilon) B\}$. In order to simplify the notation, given a set $x$ we define $\skewm_i(\epsilon, x)$ to be the event that $a^S_i(x) \le (1 - \epsilon) B$ and $\skewp_i(\epsilon, x)$ to be the event that $a^S_i(x) \ge (1 - 2\epsilon) B$. Notice that if $x \in \mathcal{X}_i^+$, then $\skewm_i(\epsilon, x)$ is the event that $a_i^S(x)$ is significantly smaller than its expectation (skewed in the minus direction), while for $x \in \mathcal{X}_i^-$ $\skewp_i(\epsilon, x)$ is the event that $a_i^S(x)$ is significantly larger than its expectation (skewed in the plus direction). These definitions directly give the equivalence
	%
	\begin{equation*} \label{eq:witness1}
		\Pr\left(\bigvee_{i,x \in \mathcal{X}} x \textrm{ can be badly learned for budget } i\right) = \Pr\left(\bigvee_{i, x \in \mathcal{X}_i^+} \skewm_i(\epsilon, x) \vee \bigvee_{i, x \in \mathcal{X}_i^-} \skewp_i(\epsilon, x)\right).
	\end{equation*}
	
	In order to introduce the concept of witnesses, consider two sets $x,x'$, say, in $\mathcal{X}_i^+$. Take a subset $w \subseteq x \cap x'$; the main observation is that, since $a^t \ge 0$ for all $t$, for all scenarios we have $a_i^S(w) \le a_i^S(x)$ and $a_i^S(w) \le a_i^S(x')$. In particular, the event $\skewm_i(\epsilon, x) \vee \skewm_i(\epsilon,x')$ is contained in $\skewm(\epsilon, w)$. The set $w$ serves as a witness for scenarios which are skewed for either $x$ or $x'$; if additionally $a_i(w)$ reasonably larger than $(1 - \epsilon) B$, we can then use concentration inequalities over $\skewm_i(\epsilon, w)$ in order to bound probability of $\skewm(\epsilon, x) \vee \skewm(\epsilon,x')$. This ability of bounding multiple terms of the right-hand side of \eqref{eq:witness1} simultaneously is what gives an improvement over the naive union bound.
	
	\balemma
	\begin{definition}[Witness] \label{def:witness}
	 We say that $\mathcal{W}_i^+$ is a \emph{witness set} for $\mathcal{X}_i^+$ if: (i) for all $w \in \mathcal{W}_i^+$, $a_i(w) \ge (1 - \epsilon/2) B$ and (ii) for all $x \in \mathcal{X}_i^+$ there is $w \in \mathcal{W}_i^+$ contained in $x$. Similarly, we say that $\mathcal{W}_i^-$ is a \emph{witness set} for $\mathcal{X}_i^-$ if: (i) for all $w \in \mathcal{W}_i^-$, $a_i(w) \le (1 - 3 \epsilon/2) B$ and (ii) for all $x \in \mathcal{X}_i^-$ there is $w \in \mathcal{W}_i^-$ containing $x$.
	\end{definition}
	\balemma

	As indicated by the previous discussion, given witness sets $\mathcal{W}_i^+$ and $\mathcal{W}_i^-$ for $\mathcal{X}_i^+$ and $\mathcal{X}_i^-$, we directly get the bound
	\begin{align} \label{eq:witness2}
		\Pr\left(\bigvee_{i, x \in \mathcal{X}_i^+} \skewm(\epsilon, x) \vee \bigvee_{i, x \in \mathcal{X}_i^-} \skewp(\epsilon, x)\right) \le \Pr\left(\bigvee_{i, w \in \mathcal{W}_i^+} \skewm(\epsilon, w) \vee \bigvee_{i, w \in \mathcal{W}_i^-} \skewp(\epsilon, w)\right).
	\end{align}
	Putting together the last three displayed equations and using Chernoff-type bounds, we can get an upper estimate on the probability that $x^S$ is bad in terms of the size of witnesses sets.
	
	\balemma
	\begin{lemma} \label{lemma:badWitness}
		Suppose that, for all $i \in [m]$, there are witness sets for $\mathcal{X}_i^+$ and $\mathcal{X}_i^-$ of size at most $M$. Then $\Pr(x^S \textrm{ is bad }) \le 8mM \exp\left(-\frac{\epsilon^3 B}{33}\right)$.
	\end{lemma}
	\balemma
	
	One natural choice of a witness set for, say, $\mathcal{X}_i^+$ is the collection of all of its minimal sets; unfortunately this may not give a witness set of small enough size. But notice that a witness set need not be a subset of $\mathcal{X}_i^+$ (or even $\mathcal{X}$). Allowing elements outside $\mathcal{X}_i^+$ gives the flexibility of obtaining witnesses which are associated to multiple ``similar'' minimal elements of $\mathcal{X}_i^+$, which is effective in reducing the size of witness sets.
	
	
	\beforesubsection
	\subsection{Small witness sets for almost 1-dim columns} \label{sec:goodWitness}
	\aftersection
	
\begin{figure}
	\centering
		\includegraphics[scale=0.5]{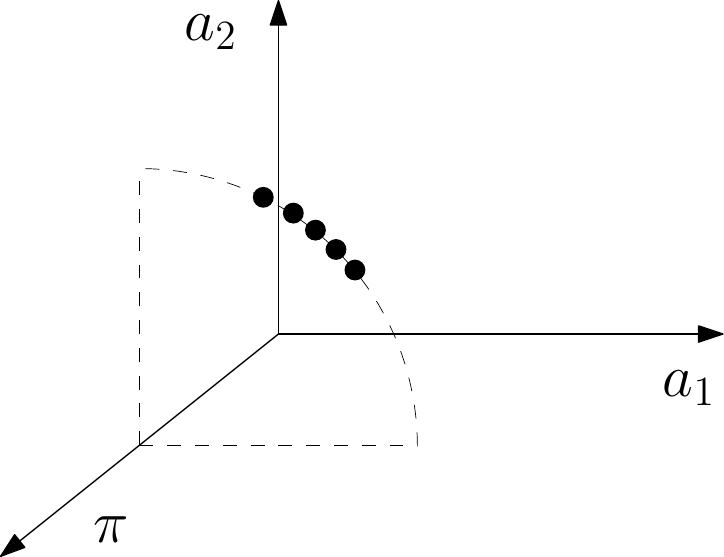}
		\caption{Case $m = 2$, columns $(\pi_t, a^t)$ equal to $(1,\sin(\frac{\pi}{4} + \delta t), \cos(\frac{\pi}{4} + \delta t))$ for sufficiently small $\delta > 0$, represented by black dots. Each segment $\{t, t + 1, \ldots, t + j\}$ can be linearly classified and hence belongs to $\mathcal{X}$. Furthermore, all segments $\{j 2 B, \ldots, (j + 1) 2 B\}$ belong to $\mathcal{X}_i^+$, which then contains $\Omega(\frac{n}{B})$ disjoint sets. Similar analysis holds for $\mathcal{X}_i^-$.}
	\label{fig:disjoint3d}
\end{figure}

Given the previous lemma, our task is to find small witness sets. Unfortunately, when the $(\pi_t, a^t)$'s lie in a space of dimension at least 3, $\mathcal{X}_i^+$ and $\mathcal{X}_i^-$ may contain many ($\Omega(n)$) disjoint sets (see Figure \ref{fig:disjoint3d}), which shows that in general we cannot find small witness sets directly. This sharply contrasts with the case where the $(\pi_t, a^t)$'s lie in a 2-dimensional subspace of $\mathbb{R}^{m + 1}$, where one can show that $\mathcal{X}$ is a union of 2 chains with respect to inclusion. In the special case where the $a^t$'s lie in a 1-dimensional subspace of $\mathbb{R}^m$, we show that $\mathcal{X}$ is actually a single chain (Lemma \ref{lemma:chain}) and therefore we can take $\mathcal{W}_i^+$ as \emph{the} minimal set of $\mathcal{X}_i^+$ and $\mathcal{W}_i^-$ as \emph{the} maximal set of $\mathcal{X}_i^-$.

	Due to the above observations, we focus on LP's whose $a^t$'s lie in few 1-dimensional subspaces. In this case, $\mathcal{X}_i^+$ and $\mathcal{X}_i^-$ are sufficiently well-behaved so that we can find small (independent of $n$) witness sets.
	
	\balemma
	\begin{lemma} \label{lemma:witness2Dim}
		Suppose that there are $K \ge m$ 1-dimensional subspaces of $\mathbb{R}^m$ which contain the $a^t$'s. Then there are witness sets for $\mathcal{X}_i^+$ and $\mathcal{X}_i^-$ of size at most $(O(\frac{K}{\epsilon} \log \frac{K}{\epsilon}))^m$.
	\end{lemma}
	\balemma
	
	Assuming the hypothesis of the lemma, partition the index set $[n]$ into $C_1, C_2, \ldots, C_K$ such that for all $j \in [K]$ the columns $\{a^t\}_{t \in C_j}$ belong to the same 1-dimensional subspace. Equivalently, for each $j \in [K]$ there is a vector $c^j$ of $\ell_\infty$-norm 1 such that for all $t \in C_j$ we have $a^t = \|a^t\|_{\infty} c^j$. An important observation is that now we can order the columns (locally) by the ratio of profit over budget occupation: without loss of generality assume that for all $j \in [K]$ and $t, t' \in C_j$ with $t < t'$, we have $\frac{\pi_t}{\|a^t\|_{\infty}} \ge \frac{\pi_{t'}}{\|a^{t'}\|}_{\infty}$.\footnote{Notice that this ratio is well-defined since by assumption $a^t \neq 0$ for all $t \in [n]$.}
	
	Given a classification $x$, we use $x|_{C_j}$ to denote its projection onto the coordinates in $C_j$; so $x|_{C_j}$ is the induced classification on columns with indices in $C_j$. 
	Similarly, we define $\mathcal{X}|_{C_j} = \{x|_{C_j} : x \in \mathcal{X}\}$ as the set of all classifications induced in the columns in $C_j$. The most important structure that we get from working with 1-d subspaces, which is implied by the local order of the columns, is the following.
	
	\balemma
	\begin{lemma} \label{lemma:chain}
		For each $j \in [K]$, the sets in $\mathcal{X}|_{C_j}$ are prefixes of $C_j$.
	\end{lemma}
	\balemma
	
	To simplify the notation fix $i \in [m]$ for the rest of this section, so we aim at providing witness sets for $\mathcal{X}_i^+$ and  $\mathcal{X}_i^-$. 
The idea is to group the classifications according to their budget occupation caused by the different column classes $C_j$'s. To make this formal, 
start by covering the interval $[0, B + m]$ with intervals $\{I_\ell\}_{\ell \in L}$, where $I_0 = [0, \frac{\epsilon B}{4K})$ and $I_\ell = [\frac{\epsilon B}{4K} (1 + \frac{\epsilon}{4})^{\ell - 1}, \frac{\epsilon B}{4K} (1 + \frac{\epsilon}{4})^\ell)$ for $\ell > 0$ and $L = \{0, \ldots, \lceil \log_{1 + \epsilon/4} \frac{8K}{\epsilon} \rceil\}$ (note that since $B \geq m$, we have $B + m \leq 2B$).
Define $\mathcal{B}_{i,j}^\ell$ as the set of partial classifications $y \in \mathcal{X}|_{C_j}$ whose budget occupation $a_i(y)$ lies in the interval $I_\ell$. For $\boldl \in L^K$ define the family of classifications 
$\mathcal{B}_i^{\boldl} = \{(y^1, y^2, \ldots, y^K) : y^j \in \mathcal{B}_{i,j}^{\boldl_j}\}$. 
The $\mathcal{B}_i^{\boldl}$'s then provide the desired grouping of the classifications.
Note that the $\mathcal{B}_i^{\boldl}$'s may include classifications not in $\mathcal{X}$ and may not include classifications in $\mathcal{X}$ which have occupation $a_i(.)$ greater than $B + m$.

	Now consider a non-empty $\mathcal{B}_i^{\boldl}$. Let $\underline{w}_i^{\boldl}$ be the inclusion-wise smallest element in $\mathcal{B}_i^{\boldl}$. Notice that such unique smallest element exists: since $\mathcal{X}|_{C_j}$ is a chain, so is $\mathcal{B}_{i,j}^{\boldl_j}$, and hence $\underline{w}_i^{\boldl}$ is the product (over $j$) of the smallest elements in the sets $\{\mathcal{B}_{i,j}^{\boldl_j}\}_j$. Similarly, let $\overline{w}_i^{\boldl}$ denote the largest element in $\mathcal{B}_i^{\boldl}$. Intuitively, $\underline{w}^{\boldl}_i$ and $\overline{w}^{\boldl}_i$ will serve as witnesses for all the sets in $\mathcal{B}_i^{\boldl}$.
	
	Finally, define the witness sets by adding the $\underline{w}_i^{\boldl}$ and $\overline{w}_i^{\boldl}$'s of appropriate size corresponding to meaningful $\mathcal{B}_i^{\boldl}$'s: set $\mathcal{W}_i^+ = \{\underline{w}_i^{\boldl} : \boldl \in L^K, \mathcal{B}_i^{\boldl} \cap \mathcal{X} \neq \emptyset, a_i(\underline{w}_i^{\boldl}) \ge (1 - \epsilon/2) B\}$ and $\mathcal{W}_i^- = \{\overline{w}_i^{\boldl} : \boldl \in L^K, \mathcal{B}_i^{\boldl} \cap \mathcal{X} \neq \emptyset, a_i(\overline{w}_i^{\boldl}) \le (1 - 3\epsilon/2) B\}$.

	It is not too difficult to see that, say, $\mathcal{W}_i^+$ is a witness set for $\mathcal{X}_i^+$: If $x \in \mathcal{X}_i^+$ belongs to some $\mathcal{B}_i^{\boldl}$, then $\underline{w}_i^{\boldl}$ belongs to $\mathcal{W}_i^+$ and is easily shown to be a witness for $x$. However, if $x$ does not belong to any $\mathcal{B}_i^{\boldl}$, by having too large $a_i(x)$, the idea is to find $x' \subseteq x$ which belongs to some $\mathcal{B}_i^{\boldl}$ \emph{and} to $\mathcal{X}$, and then use $\underline{w}_i^{\boldl}$ as a witness for $x$. We note that considering $B_i^{\boldl}$'s for side lengths at most $B + m$ and only adding witnesses for $B_i^{\boldl}$'s which intersect $\mathcal{X}$ are crucially used for bounding the size of $\mathcal{W}_i^+$ and $\mathcal{W}_i^-$.
	
	\balemma
	\begin{lemma} \label{lemma:wWitness}
		The sets $\mathcal{W}_i^+$ and $\mathcal{W}_i^-$ are witness sets for $\mathcal{X}_i^+$ and $\mathcal{X}_i^-$.
	\end{lemma}
	\balemma
	

	\myparagraph{Bounding the size of witness sets.} Clearly the witness sets $\mathcal{W}_i^+$ and $\mathcal{W}_i^-$ have size at most $|L|^K$. Although this size is independent of $n$, it is still unnecessarily large since it only uses locally (for each $C_j$) the fact that $\mathcal{X}$ consists of linear classifications; in particular, it does not use the dimension of the ambient space $\mathbb{R}^m$. Now we sketch the argument for an improved bound, and details are provided in the appendix. 
	
	First notice that the partial classification $x(p)|_{C_j}$ is completely defined by the value $pc^j$. Thus, if $J \subseteq [K]$ is such that the directions $\{c^j\}_{j \in J}$ form a basis of $\mathbb{R}^m$ then knowing $pc^j$ for all $j \in J$ completely determines the whole classification $x(p)$. Similarly, if we know that $x(p)|_{C_j} \in \mathcal{B}_i^{\boldl_j}$ for all $j \in J$, then for each $j \notin J$ we should have fewer possible $\mathcal{B}_i^{\boldl_j}$'s where the partial classification $x(p)|_{C_j}$ can belong to; this indicates that some of the sets $\{\mathcal{B}_i^{\boldl}\}_{\boldl \in L^K}$ do not contain any element from $\mathcal{X}$, which implies a reduced size for the witness sets. 
	
	In order to capture this idea, we focus on the space of dual vectors $p$ and define the sets $P^\ell_j = \{p \in \R_+^m : x(p)|_{C_j} \in \mathcal{B}_{i,j}^{\ell}\}$ 
	 and $P^v = \{p \in \R_+^m : x(p) \in \mathcal{B}_i^v\}$.
	Notice that $P^v = \cap_j P^{v_j}_j$ and that $\mathcal{B}^v_i$ is empty iff $P^v$ is. The main step is to show that each $P^\ell_j$ is a polyhedron with ``few'' facets, which uses the definition of $x(p)$ and Lemma \ref{lemma:chain}. We then consider the arrangement of the hyperplanes which are facet-defining for the $P_j^\ell$'s and conclude that the $P^v$'s are given by unions of the cells in this arrangement; classical bounds on the number of cells in a hyperplane arrangement in $\R^m$ then allow us to upper bound the number of nonempty $P^v$'s. This gives the following.
	
	\balemma
	\begin{lemma} \label{lemma:sizeP}
		At most $(O(\frac{K}{\epsilon} \log \frac{K}{\epsilon}))^m$ of the $\mathcal{B}_i^{\boldl}$'s contain an element from $\mathcal{X}$.
	\end{lemma}
	\balemma
	
%
%
%

	This lemma implies that $\mathcal{W}_i^+$ and $\mathcal{W}_i^-$ each has size at most $(O(\frac{K}{\epsilon} \log \frac{K}{\epsilon}))^m$, which then proves Lemma \ref{lemma:witness2Dim}.	Finally, applying Lemma \ref{lemma:badWitness} we conclude the proof of Lemma \ref{lemma:goodOtp}.

	\beforesection
	\section{Robust \otp} \label{sec:robustOTP}
	\aftersection

\label{sec:rotp}	
	In this section we consider \eqref{eq:LP} with columns that may not belong to few 1-dimensional subspaces. Given the results of the previous section we would like to perturb the columns of this LP so that it belongs to few 1-dim subspaces, and such that an approximate solution for this perturbed LP is also an approximate solution for the original one. More precisely, we obtain a set of vectors $Q \subseteq \mathbb{R}^m$ and 
transform each column $a^t$ into a column $\tilde{a}^t$ which is a scaling of a vector in $Q$, 
and we let the rewards $\pi_t$ remain unchanged. The crucial observation is that the solutions of an LP are robust to slight changes in the the constraint matrix.
	
	\balemma
	\begin{lemma} \label{lemma:robust}
		Consider real numbers $\pi_1, \ldots, \pi_n$ and vectors $a^1, \ldots, a^n$ and $\tilde{a}^1, \ldots, \tilde{a}^n$ in $\mathbb{R}^m_+$ such that $\|\tilde{a}^t - a^t\|_{\infty} \le \frac{\epsilon}{m + 1} \|a^t\|_{\infty}$. If $x$ is an $\epsilon$-approximate solution for \eqref{eq:LP} with columns $(\pi_t, \tilde{a}^t)$ and right-hand side $(1 - \epsilon) B$, then $x$ is a $2\epsilon$-approximate solution for the LP \eqref{eq:LP}.
	\end{lemma}
	\balemma
	
	\myparagraph{Perturbing the columns.} To simplify the notation, set $\delta = \frac{\epsilon}{m+1}$; for simplicity of exposition we assume that $1/\delta$ is integral. When constructing $Q$ we want the rays spanned by the each of its vectors to be ``uniform'' over $\mathbb{R}^m_+$. Using $\ell_\infty$ as normalization, let $Q$ be a $\delta$-net of the unit $\ell_\infty$ sphere, namely let $Q$ be the vectors in $\{0, \delta, 2\delta, 3\delta, \ldots, 1\}^m$ which have $\ell_\infty$ norm 1. Note that $|Q| = (O(\frac{m}{\epsilon}))^m$.
	
	
	Given a vector $a^t \in \R^m$ 
we let $\tilde{a}^t = \|a^t\|_\infty q^t$, where $q^t$ is the vector in $Q$ closest (in $\ell_\infty$) to $\frac{a^t}{\|a^t\|_\infty}$. By definition of $Q$, for every vector $v \in \mathbb{R}^m$ with $\|v\|_\infty = 1$ there is a vector $q \in Q$ with $\|v - q\|_{\infty} \le \delta$. It then follows from positive homogeneity of norms that the $\tilde{a}^t$'s satisfy the property required in Lemma \ref{lemma:robust}: 
$\|a^t - \tilde{a}^t\|_{\infty} \le \delta \|a^t\|_{\infty}$. 
	
	\myparagraph{Algorithm Robust \otp.} One way to think of the algorithm Robust \otp is that it works in two phases. First, it transforms the vectors $a^t$ into $\tilde{a}^t$ as described above. Then it returns the solution obtained by running the algorithm \otp over the LP with columns $(\pi_t, \tilde{a}^t)$ and right-hand side $(1 - \epsilon)B$. Notice that this algorithm can indeed be implemented to run in an online fashion.
	
	Putting together the discussion in the previous paragraphs and the guarantee of \otp for almost 1-dim columns given by Theorem \ref{thm:otp} with $K = |Q| = (O(\frac{m}{\epsilon}))^m$, we obtain the following theorem.
	
	\balemma
	\begin{theorem} \label{thm:rotp}
		Fix $\epsilon \in (0,1]$ and suppose $B \ge \Omega\left(\frac{m^2}{\epsilon^3} \log \frac{m}{\epsilon}\right)$. Then algorithm Robust \otp returns a solution to the online \eqref{eq:LP} with expected value at least $(1 - 10\epsilon) \OPT$.
	\end{theorem}
	\balemma


\beforesection
\section{Robust \dpa} \label{sec:rdpa}
\aftersection

	In this section we describe our final algorithm, which has an improved dependence on $1/\epsilon$. Following \cite{agrawal}, the idea is to update the dual vector used in the classification as new columns arrive: we use the first $2^i \epsilon n$ columns to classify columns $2^i \epsilon n + 1, \ldots, 2^{i + 1} \epsilon n$. This leads to improved generalization bounds, which in turn give the reduced dependence on $1/\epsilon$. The algorithm Robust \dpa (as the algorithm \dpa) can be seen as a combination of solutions to multiple sampled LP's, obtained via a modification of \otp denoted by $(s,\delta)$-\otp.
	
	\myparagraph{Algorithm $(s,\delta)$-\otp.} This algorithm aims at solving the program $(2s,1)$-LP and can be described as follows: it finds an optimal dual solution $(p, \alpha)$ for $(s, (1 - \delta))$-LP and sets $x_{\sigma(t)} = x(p)_{\sigma(t)}$ for $t = s + 1, s + 2, \ldots, t' \le 2s$ such that $t'$ is the maximum one guaranteeing $\sum_{t = s + 1}^{2s} a^{\sigma(t)} x_{\sigma(t)} \le \frac{s}{n} B$.
	
	The analysis of $(s,\delta)$-\otp is similar to the one employed for \otp. The main difference is that this algorithm tries to approximate the value of the \emph{random} LP $(2s,1)$-LP. This requires a partition of the bad classifications which is more refined than simply splitting into $\mathcal{X}_i^+$ and $\mathcal{X}_i^-$, and witness sets need to be redefined appropriately. Nonetheless, using these ideas we can prove the following guarantee for $(s,\delta)$-\otp. Again let $S = \{\sigma(1), \sigma(2), \ldots, \sigma(s)\}$ be the random index set of the first $s$ columns of the LP, let $T = \{\sigma(s+1), \sigma(s+2), \ldots, \sigma(2s)\}$ and $U = S \cup T$. We use $\pi_U$ to denote the vector $(\pi_t)_{t \in U}$. 
	
	\balemma
	\begin{lemma} \label{lemma:sDeltaOtp}
		Suppose that there are $K \ge m$ 1-dim subspaces of $\mathbb{R}^m$ containing the columns $a^t$'s. Fix an integer $s$ and a real number $\delta \in (0,1/10)$ such that $\frac{\delta^2 s B}{n} \ge \Omega(m \ln \frac{K}{\delta})$. Then algorithm $(s,\delta)$-\otp returns a solution $x$ satisfying $a_i^T(x) \le B$ for all $i \in [m]$ with probability 1 and with expected value $\E[\pi_U x] \ge (1 - 3 \delta) \E[\OPT(2s)] - \E[\OPT(s)] - \delta^2 \OPT$.
	\end{lemma}	 
	\balemma
	

	\myparagraph{Algorithm Robust \dpa.} In order to simplify the description of the algorithm, we assume in this section that $\log (1/\epsilon)$ is an integer.

	Again the algorithm Robust \dpa can be thought as acting in two phases. In the first phase it converts the vectors $a^t$ into $\tilde{a}^t$, just as in the first phase of Robust \otp. In the second phase, for $i = 0, \ldots, \log (1/\epsilon) - 1$, it runs $(\epsilon 2^i n, \sqrt{\epsilon/2^i})$-\otp over \eqref{eq:LP} with columns $(\pi_t, \tilde{a}^t)$ and right-hand side $(1 - \epsilon) B$ to obtain the solution $x^i$. The algorithm finally returns the solution $x$ consisting of the ``union'' of $x^i$'s: $x = \sum_i x^i$.
	
	Note that the second phase corresponds exactly to using the first $\epsilon 2^i n$ columns to classify the columns $\epsilon 2^i n + 1, \ldots, \epsilon 2^{i+1} n$. This relative increase in the size of the training data for each learning problem allow us to reduce the dependence of $B$ on $\epsilon$ in each of the iterations, while the error from all the iterations telescope and are still bounded as before. Furthermore, notice that Robust \dpa can be implemented to run online.
		
	The analysis of Robust \dpa reduces to that of $(s,\delta)$-\otp. That is, using the definition of the parameters of $(s,\delta)$-\otp used in Robust \dpa and Lemma \ref{lemma:sDeltaOtp}, it is routine to check that the algorithm produces a feasible solution which has expected value $(1 - \epsilon) \OPT$. This is formally stated in the following theorem.
		
	\balemma
	\begin{theorem} \label{thm:expValueDPA}
		Fix $\epsilon \in (0,1/100)$ and suppose that $B \ge \Omega(\frac{m^2}{\epsilon^2} \ln \frac{m}{\epsilon})$. Then the algorithm Robust \dpa returns a solution to the online LP \eqref{eq:LP} with expected value at least $(1 - 50\epsilon) \OPT$.
	\end{theorem}
	\balemma


	\beforesection
	\section{Open problems}
	\aftersection
	
	A very interesting open question is whether the techniques introduced in this work can be used to obtain improved algorithms for generalized allocation problems \cite{feldman}. The difficulty in this problem is that the classifications of the columns are not linear anymore; they essentially come from a conjunction of linear classifiers. Given this additional flexibility, having the columns in few 1-dimensional subspaces does not seem to impose strong enough properties in the classifications. It would be interesting to find the appropriate geometric structure of the columns in this case.
	
	Of course a direct open question is to improve the lower or upper bound on the dependence on the right-hand side $B$ to obtain $(1 - \epsilon)$-competitive algorithms. One possibility is to investigate how much the techniques presented here can be pushed and what are their limitations. Another possibility is to analyze the performance of the algorithm from \cite{Devanur11} under the random permutation model.  

	\bibliographystyle{abbrv}
	\bibliography{online-lp}

	\pagebreak
	
	\appendix
	
	\section{Bernstein inequality for sampling without replacement}
	
	\begin{lemma}[Theorem 2.14.19 in \cite{weakConvergence}] \label{lemma:addChernoff}
		Let $Y = \{Y_1, \ldots, Y_n\}$ be a set of real numbers in the interval $[0,1]$ and let $0 < \epsilon < 1$. Let $S$ be a random subset of $Y$ of size $s$ and let $Y_S = \sum_{i \in S} Y_i$. Setting $\mu = \frac{1}{n} \sum_i Y_i$ and $\sigma^2 = \frac{1}{n}\sum_i (Y_i - \mu)^2$, we have that for every $\tau > 0$
		\begin{align*}
			\Pr(|Y_S - s \mu| \ge \tau) \le 2 \exp \left( -\frac{\tau^2}{2 s \sigma^2 + \tau} \right)
		\end{align*}
	\end{lemma}
	
	Notice that, since the $Y_i$'s belong to the interval $[0,1]$, we can upper bound the variance by the mean as follows: $$\sigma^2 \le \frac{1}{n} \sum_i |Y_i - \mu| \le \frac{1}{n} \left( \sum_i |Y_i| + \sum_i |\mu| \right) = 2 \mu.$$ This gives the following corollary.
		
	\begin{corollary} \label{cor:multiChernoff}
		Consider the conditions of the previous lemma. Then for all $\tau > 0$
		\begin{align*}
			\Pr(|Y_S - s \mu| \ge \tau) \le 2 \exp \left( -\frac{\tau^2}{4 s \mu + \tau} \right).
		\end{align*}
	\end{corollary}
	
	
	\section{Proof of Lemmas \ref{lemma:approximateCS} and \ref{lemma:sampleCS}}

	\begin{proofof}{Lemma \ref{lemma:approximateCS}}
		Fix a scenario $\sigma$ for the duration of the proof. By assumption $x^S$ is feasible for \eqref{eq:LP}, so it suffices to show that it attains value at least $(1-3\epsilon) \OPT$. For that, consider \eqref{eq:LP} with a modified right-hand side:
		\begin{align}
			\max \sum_{t = 1}^n \pi_t x_t \notag\\
			\sum_{t = 1}^n a^t_i x_t \le a_i(x^S)  \ \ \ \ \forall i \in [m] \tag{modLP} \label{eq:modLP} \\
			x \in [0,1]^n. \notag
		\end{align}
		Consider the Lagrangian relaxation $L(p, x) = \sum_{t = 1}^{\epsilon n} \pi_t x_t - \sum_{i = 1}^m p_i (\sum_{t = 1}^{\epsilon n} a^t_i x_t - a_i(x^S))$. Notice that $x^S$ is an optimal solution for $\max_{x \in [0,1]^n} L(p^S, x)$, which is at least the \OPT(\ref{eq:modLP}), the optimum value of LP \eqref{eq:modLP}. Since $x^S$ is clearly feasible for \eqref{eq:modLP}, it follows that $x^S$ is an optimal solution for the latter.
		
		Now let $x^*$ be an optimal solution for \eqref{eq:LP}. Since $a_i(x^S) \ge (1 - 3 \epsilon) B$ for all $i$, and since $a^t \ge 0$ for all $t$, it follows that $(1 - 3 \epsilon) x^*$ is feasible for \eqref{eq:modLP}. By linearity of the objective function we get that \OPT(\ref{eq:modLP}) $ \ge (1 - 3\epsilon) \sum_{t = 1}^n \pi_t x^*_t  = (1 - 3 \epsilon) \OPT$ and the result follows.
	\end{proofof}
	
	
	\medskip
	\begin{proofof}{Lemma \ref{lemma:sampleCS}}
	 Fix a scenario $\sigma$ for the duration of the proof. Let $x^*$ be an optimal solution for $(\epsilon n, (1 - \epsilon))$-LP in complementary slackness with $p^S$. If $p^S a^t > \pi_t$, the corresponding constraint in the dual is loose and by complementary slackness we get $x^*_t = 0$. If $p^S a^t < \pi_t$, then for dual feasibility we have $\alpha^*_t > 0$ and by complementary slackness we have $x^*_t = 1$.
		
	 From the definition of $x^S$ we get that $x^S \le x^*$ and, since the $a^t$'s are non-negative, the feasibility of $x^*$ implies that $a_i^S(x^S) \le (1-\epsilon)B$ for all $i \in [m]$. Moreover, from our assumption that the input is in general position we get that there are at most $m$ values of $t$ such that $p^S a^t = \pi_t$. Therefore, $x^S$ and $x^*$ differ in at most $m$ positions and from primal complementary slackness we get that whenever $p^S > 0$, $a_i^S(x^S) \ge a_i^S(x^*) - m = (1 - \epsilon) B - m \ge (1 - 2\epsilon) B$, where the last inequality follows from the fact that $B \ge \frac{1}{\epsilon}$. This concludes the proof of the lemma.
	\end{proofof}


	\section{Proof of Lemma \ref{lemma:badWitness}}
	
	The following simple inequalities will be helpful.
	
	\begin{observation} \label{obs:ratio}
		For $\epsilon, \alpha, \beta \ge 0$, $\frac{1-\alpha \epsilon}{1 + \beta \epsilon} \ge 1 - (\alpha + \beta) \epsilon$ and $\frac{1-\alpha \epsilon}{1 - \beta \epsilon} \le 1 - (\alpha - \beta) \epsilon$.
	\end{observation}
	
	Combining equations \eqref{eq:badlyLearned}, \eqref{eq:witness1} and \eqref{eq:witness2} and union bounding over all terms in the disjunction, we have that
	\begin{align*}
		\Pr\left(x^S \textrm{ is bad}\right) \le \sum_{i, w \in \mathcal{W}_i^+} \Pr\left(\skewm(\epsilon,w)\right) + \sum_{i, w \in \mathcal{W}_i^-} \Pr\left(\skewp(\epsilon,w)\right).
	\end{align*}
	Thus, it suffices to show that for all $w \in \mathcal{W}_i^+$ (respectively $w \in \mathcal{W}_i^-$), the event $\skewm(\epsilon, w)$ (resp. $\skewp(\epsilon, w)$) occurs with probability at most $2\exp\left(-\frac{\epsilon^3 B}{33}\right)$.
	
	Take $w \in \mathcal{W}_i^+$. By definition of this set, $a_i(w) \ge (1 - \frac{\epsilon}{2})B$, so the event $\skewm(\epsilon,w)$ is contained in the event that $a_i^S(w) \le (1 - \epsilon) a_i(w)/(1 - \frac{\epsilon}{2})$, which is contained in the event $a_i^S(w) \le (1 - \frac{\epsilon}{2}) a_i(w)$. Using Corollary \ref{cor:multiChernoff} with $\tau = \epsilon^2 a_i(w) /2$, we obtain that $\Pr(\skewm(\epsilon, w)) \le 2\exp\left(-\frac{\epsilon^3 B}{33}\right)$.
	
	Similarly, take $w \in \mathcal{W}_i^-$, such that $a_i(w) \le (1 - \frac{3\epsilon}{2})B$. It is easy to check that the event $\skewp(\epsilon,w)$ is contained in $a_i^S(w) \ge (1 + \frac{\epsilon}{2})a_i(w)$, so using Corollary \ref{cor:multiChernoff} with $\tau = \epsilon^2 B/2$ we get that  $\Pr(\skewm(\epsilon, w)) \le 2\exp\left(-\frac{\epsilon^3 B}{33}\right)$. This concludes the proof of the lemma.
		

	\section{Proof of Lemma \ref{lemma:chain}}

		Fix $j \in [K]$. Consider a set $x \in \mathcal{X}$ and let $p$ be a dual vector such that $x(p) = x$. Let $t'$ be the last index of $C_j$ which belongs to $x|_{C_j}$; this implies that $\pi_{t'} > p a^{t'} = p c^j \|a^{t'}\|_{\infty}$, or alternatively $\frac{\pi_{t'}}{\|a^{t'}\|_{\infty}} > p c^j$. By the ordering of the columns, for all $t \in C_j$ smaller than $t'$ we have $\frac{\pi_t}{\|a^t\|_{\infty}} \ge \frac{\pi_{t'}}{\|a^{t'}\|_{\infty}} > p c^j$ and hence $t \in x|_{C_j}$. By definition of $t'$ it follows that $x|_{C_j} = \{t \in C_j : t \le t'\}$, a prefix of $C_j$; this concludes the proof.


	\section{Proof of Lemma \ref{lemma:wWitness}}

		We prove that $\mathcal{W}_i^+$ is a witness set for $\mathcal{X}_i^+$; the proof that $\mathcal{W}_i^-$ is a witness set for $\mathcal{X}_i^-$ is analogous.
	
		First, we claim that for all $x \in \mathcal{X}_i^+$, there is $x' \in \mathcal{X}$ such that $x' \subseteq x$ and $a_i(x') \in [B, B + m]$. To see this, let $p$ be such that $x = x(p)$. For $\lambda \ge 0$, define $p^\lambda = p + \lambda e_i$, where $e_i$ denotes the $i$th canonical vector. We have that $a_i(x(p^0)) > B$ (since $x(p) \in \mathcal{X}_i^+$) and $a_i(x(p^\infty)) = 0$ (since columns with $a_i^t > 0$ will at have at some point $p^\lambda a^t \ge \pi_t$). Due to the assumption that the input is in general position, whenever $a_i(x(p^\lambda))$ is discontinuous (as a function of $\lambda \ge 0$) the right and the left limits differ by at most $m$. It then follows that there is $\lambda \ge 0$ such that $a_i(x(p^\lambda)) \in [B, B + m]$, and since $x(p^\lambda) \subseteq x$ for all $\lambda \ge 0$ the claim follows.
		
		So take a classification $x \in \mathcal{X}_i^+$ and let $x'$ be as above. The fact that $a_i(x') \le B + m$ and the non-negativity of the $a^t$'s imply that there is an $\ell \in L^K$ such that $x' \in \mathcal{B}_i^\ell$. Since $\underline{w}^\ell$ is the unique smallest set in $\mathcal{B}_i^\ell$, clearly $x' \subseteq \underline{w}^\ell$. To show that $\underline{w}^\ell \in \mathcal{W}_i^+$, it suffices to argue that $a_i(\underline{w}^\ell) \ge (1 - \epsilon/2) B$.
		
		Since $\underline{w}^\ell, x' \in \mathcal{B}_i^\ell$, for all $j$ such that $\ell_j > 0$ we have $a_i(\underline{w}^\ell|_{C_j}) \ge a_i(x'|_{C_j}) / (1 + \frac{\epsilon}{4})$. Moreover, for $j$ such that $\ell = 0$ we have $a_i(x(p)|_{C_j}) < \frac{\epsilon B}{4K}$. Adding over all $j \in [K]$ gives
		\begin{align*}
			a_i(\underline{w}^\ell) \ge \left( \frac{1}{1 + \frac{\epsilon}{4}} \right) \left[ a_i(x(p)) - \sum_{j : \ell_j = 0} a_i(x(p)|_{C_j}) \right] \ge \frac{B}{1 + \frac{\epsilon}{4}} - \frac{\epsilon B}{4} \ge \left(1 - \frac{\epsilon}{2}\right) B,
		\end{align*}
		where the third inequality follows from Observation \ref{obs:ratio}. Thus, $\underline{w}^\ell \in \mathcal{W}_i^+$.
		
		 Since this property holds for all $x \in \mathcal{X}_i^+$, we conclude that $\mathcal{W}_i^+$ is a witness set for $\mathcal{X}_i^+$.
				

\section{Proof of Lemma \ref{lemma:sizeP}}

	Recall the definitions of $P^{\boldl}$ (for $\boldl \in L^K$) and $P_j^{\ell}$ (for $j \in [m]$, $\ell \in L$). It suffices to prove that at most $(O(\frac{K}{\epsilon} \log \frac{K}{\epsilon}))^m$ of the families $P^{\boldl}$'s are non-empty.
		
		Since $x(p) \in \mathcal{B}_i^{\boldl}$ if and only if for all $j \in [K]$ we have $x(p)|_{C_j} \in \mathcal{B}_{i,j}^{\boldl_j}$, it follows that $P^{\boldl} = \bigcap_j P_j^{\boldl_j}$. Let $\tau^\ell_j$ denote the first index in $C_j$ such that the prefix $\{t \in C_j : t \le \tau^\ell_j\}$ occupies the budget $i$ to an extent in $I_\ell$. Using Lemma \ref{lemma:chain} and the fact that the $a^t$'s are non-negative, we get that $\mathcal{B}_{i,j}^\ell$ is the set of all prefixes of $C_j$ which contain $\tau_j^\ell$ but do not contain $\tau_j^{\ell + 1}$. Moreover, notice that the set $x(p)|_{C_j}$ contains $\tau_j^\ell$ if and only if $\pi_{\tau_j^\ell} > p a^{\tau_j^\ell}$. It then follows from these observations we can express the set $P_j^\ell$ using linear inequalities: $P_j^\ell = \{p \in \mathbb{R}^m_+ : \pi_{\tau_j^\ell} > p a^{\tau_j^\ell}, \pi_{\tau_j^{\ell + 1}} \le p a^{\tau_j^{\ell + 1}}\}$. Since $P^{\boldl} = \bigcap_j P_j^{\boldl_j}$, we have that $P^{\boldl}$ is given by the intersection of halfspaces defined by hyperplanes of the form $\pi_{\tau_j^\ell} = p a^{\tau_j^\ell}$ and $p_k = 0$ ($k \in [m]$).
		 		
		So consider the arrangement given by all hyperplanes $\{\pi_{\tau_j^\ell} = p a^{\tau_j^\ell}\}_{j \in [K], \ell \in L}$ and $\{p_i = 0\}_{i = 1}^m$. Given a face $F$ in this arrangement and a set $P^{\boldl}$, either $F$ is contained in $P^{\boldl}$ or these sets are disjoint. Since the faces of the arrangement cover $\mathbb{R}^m$, it follows that each non-empty $P^{\boldl}$ contains at least one of these faces.
		
		Notice that the arrangement is defined by $K|L| + m \le O(\frac{Km}{\epsilon} \log \frac{K}{\epsilon})$ hyperplanes, where the last inequality uses the fact that $\log (1 + \frac{\epsilon}{4}) \ge \epsilon \log (1 + \frac{1}{4})$ holds (by concavity) for $\epsilon \in [0,1]$. It is known that an arrangement with $h \ge m$ hyperplanes in $\mathbb{R}^m$ has at most $\left(\frac{e h}{m}\right)^m$ faces (see Section 6.1 of \cite{matousek} and page 82 of \cite{matousekNesetril}). Using the conclusion of the previous paragraph, we get that there are at most $(O(\frac{K}{\epsilon} \log \frac{K}{\epsilon}))^m$ non-empty $P^{\boldl}$'s and the result follows.
	
				
	\section{Proof of Lemma \ref{lemma:robust}}
	
	Let LP1 denote the LP with columns $(\pi_t, \tilde{a}^t)$ and right-hand side $(1-\epsilon)B$ and LP2 denote the LP with columns $(\pi_t, a^t)$ and right-hand side $B$.
	
	Let $x$ be an $\epsilon$-approximate solution for LP1. Notice that we can upper bound $\|a^t - \tilde{a}^t\|_{\infty}$ as a function of $\|\tilde{a}^t\|_{\infty}$:
	\begin{align*}
		\|\tilde{a}^t\|_{\infty} \ge \|a^t\|_{\infty} - \|a^t - \tilde{a}^t\|_{\infty} \ge \frac{m}{\epsilon} \|a^t - \tilde{a}^t\|_{\infty},
	\end{align*}
	where the first inequality follows from triangle inequality. That is, we have $\|a^t - \tilde{a}^t\|_{\infty} \le \frac{\epsilon}{m} \|\tilde{a}^t\|_{\infty}$.
	
	Given this bound, it is easy to see that $x$ is feasible for LP2:
	\begin{align*}
		\sum_t a_i^t x_t \le \sum_t (\tilde{a}_i^t + \|a_i^t - \tilde{a}_i^t\|) x_t \le (1 - \epsilon) B + \sum_t \|a^t - \tilde{a}^t\|_{\infty} x_t \le (1 - \epsilon) B + \frac{\epsilon}{m} \sum_t \|\tilde{a}^t\|_{\infty} x_t \le B,
	\end{align*}
	where the last inequality uses the fact that $\sum_t \|\tilde{a}^t\|_{\infty} x_t \le \|\tilde{a}^t\|_1 x_t \le mB$, since $x$ is a feasible solution and the $\tilde{a}^t$'s are non-negative.
	
	In order to show that $x$ is a $2\epsilon$-approximate solution for LP2, it suffices to show that the optimum of LP1 is at least $1/(1+\epsilon)$ times the optimum of the LP2, since then $x$ will be within a factor of $(1-\epsilon)/(1+\epsilon) \ge (1 - 2\epsilon)$ the optimum of LP2. So let $x^*$ be an optimal solution for LP2. Using the same argument as before, it is easy to see that $x^*/(1+\epsilon)$ is feasible for LP1; this concludes the proof of the lemma.
	

	\section{Proof of Lemma \ref{lemma:sDeltaOtp}}

	The proof uses the same ideas used in the analysis of \otp, although some definitions need to be changed slightly.
	
	Recall that $S = \{\sigma(1), \sigma(2), \ldots, \sigma(s)\}$, $T = \{\sigma(s + 1), \sigma(s + 2), \ldots, \sigma(2s)\}$ and $U = S \cup T$. Again we use $p^S$ to denote the dual vector used by $(s, \delta)$-\otp for its classification, and set $x^S = x(p^S)$. With slight abuse in the notation, we often see $x^S$ as a (possibly infeasible) solution for $(2s,1)$-LP, which means that we truncate the vector $x^S$ to the first $2s$ coordinates $x^S_{\sigma(1)}, \ldots, x^S_{\sigma(2s)}$.

	As before, we focus on proving the following lemma; the proof that this lemma implies Lemma \ref{lemma:sDeltaOtp} is presented at the end of this section.
	
	\begin{lemma} \label{lemma:goodSDOTP}
		Suppose that there are $K \ge m$ 1-dim subspaces of $\mathbb{R}^m$ containing the columns $a^t$'s. Fix an integer $s$ and a real number $\delta \in (0,1/10)$ such that $\frac{\delta^2 s B}{n} \ge \Omega(m \ln \frac{K}{\delta})$. Then with probability at least $(1 - \delta^2)$, $x^S$ satisfies $a_i^T(x^S) \le B$ for all $i \in [m]$ and has value $\pi_U x^S \ge (1 - 3\delta) \OPT(2s)$.
	\end{lemma}
	
	In a given scenario, we now say that $x^S$ is \emph{bad} if $a_i^T(s^S) > B$ for some $i \in [m]$ or if $\pi_U x^S (1 - 3\delta) \OPT(2s)$. In this scenario, now a classification $x \in \mathcal{X}$ can be \emph{badly learned for budget $i$ due to infeasibility} if $a_i^S(x) \le (1 - \delta) B$ and $a_i^T(x) > B$; $x$ can be \emph{badly learned for budget $i$ due to value} if $a_i^S(x) \ge (1 - 2\delta)B$ and $a_i^U(x) < (1 - 3\delta)B$. Then $x$ can be \emph{badly learned for budget $i$} if it falls into any of the above cases. The following is the appropriate modification of Lemma \ref{lemma:approximateCS} for our current setting, and can be proved exactly in the same way.
	
	\begin{lemma} \label{lemma:approximateCS2}
	 	Consider a scenario where $x^S$ satisfies the following: (i) for all $i \in [m]$, $a_i^T(x^S) \le B$ and (ii) for all $i \in [m]$ with $p^S_i > 0$, $a_i^U(x^S) \ge (1 - 3 \delta) B$. Then $x^S$ is good.
	\end{lemma}
	 Due to our definitions, this lemma implies that inequality \eqref{eq:badlyLearned} still hold.
	
	\paragraph{Witness sets.} In the analysis of \otp, each $x \in \mathcal{X}$ could be badly learned for budget $i$ due to either infeasibility or (exclusively) due to value, which motivated the definitions of $\mathcal{X}_i^+$ and $\mathcal{X}_i^-$. Now the same $x$ can be badly learned for budget $i$ due to both conditions. Therefore, we introduce two different partition of $\mathcal{X}$, which tells \emph{why} a classification is unlikely to be badly learned due to the appropriate condition. That is, we define $\mathcal{X}_i^+ = \{x \in \mathcal{X} : a_i(x) > (1 - \delta) B + \frac{\delta B}{2}\}$ and $\mathcal{Y}_i^+ = \{x \in \mathcal{X} : a_i(x) \le (1 - \delta) B + \frac{\delta B}{2}\}$ as the partition associated to the infeasibility condition and $\mathcal{X}_i^- = \{x \in \mathcal{X} : a_i(x) < (1 - 2\delta) B - \frac{\delta B}{2}\}$ and $\mathcal{Y}_i^- = \{x \in \mathcal{X} : a_i(x) \ge (1 - 2\delta) B - \frac{\delta B}{2}\}$ as the partition associated to the value condition. For example, $\mathcal{X}_i^-$ is the set of classifications which are unlikely to be infeasible because of a small $a_i(.)$ value. Also, note that these classifications are all based on the total budget occupation rather than on the budget occupation in the first $2s$ columns only.
	
	Given this more refined tagging of elements in $\mathcal{X}$, we also need to redefine witness sets. We say that $(\mathcal{W}_i^+, \mathcal{W}_i^-, \mathcal{Z}_i^+, \mathcal{Z}_i^-)$ are \emph{witness sets} for $(\mathcal{X}_i^+, \mathcal{X}_i^-, \mathcal{Y}_i^-, \mathcal{Y}_i^+)$ respectively if they satisfy the following:
	\begin{align*}
		w \in \mathcal{W}_i^+ \Rightarrow a_i(w) \ge (1 - \delta) B + \frac{\delta B}{4}, x \in \mathcal{X}_i^+ \Rightarrow \exists w \in \mathcal{W}_i^+ : w \subseteq x \\
		w \in \mathcal{Z}_i^+ \Rightarrow a_i(w) \ge (1 - 2\delta) B - \frac{3\delta B}{4}, x \in \mathcal{Y}_i^- \Rightarrow \exists w \in \mathcal{Z}_i^+ : w \subseteq x \\
		w \in \mathcal{W}_i^- \Rightarrow a_i(w) \le (1 - 2\delta) B - \frac{\delta B}{4}, x \in \mathcal{X}_i^- \Rightarrow \exists w \in \mathcal{W}_i^+ : x \subseteq w \\
		w \in \mathcal{Z}_i^- \Rightarrow a_i(w) \le (1 - \delta) B + \frac{3 \delta B}{4}, x \in \mathcal{Y}_i^+ \Rightarrow \exists w \in \mathcal{W}_i^+ : x \subseteq w \, .
	\end{align*}
	
	Again to simplify the notation, given a set $x$ we define $\skewm_i^S(\delta, x)$ to be the event that $a^S_i(x) \le (1 - \delta) B$, $\skewp^S_i(\delta, x)$ to be the event that $a^S_i(x) \ge (1 - \delta) B$ and similarly replacing the set $S$ by the sets $T$ and $U$.	The following expression, which is the analogous to \eqref{eq:witness1}-\eqref{eq:witness2}, establishes the connection between the events where classifications can be badly learned and witness sets:
	\begin{align}
		\bigvee_{x \in \mathcal{X}} \{\textrm{$x$ can be badly learned for budget $i$}\} \subseteq & \left(\bigvee_{w \in \mathcal{W}_i^+} \skewm^S(\delta, w) \right) \vee \left( \bigvee_{w \in \mathcal{Z}_i^+} \skewm^{U}(3\delta, w) \right) \notag\\
		&\vee \left( \bigvee_{w \in \mathcal{W}_i^-} \skewp^S(2\delta, w) \right) \vee \left( \bigvee_{w \in \mathcal{Z}_i^-} \skewp^{T}(0, w)\right). \label{eq:witnessSDOTP}
	\end{align}
	To see that this expression holds, take $x \in \mathcal{X}$. Suppose that $x \in \mathcal{X}_i^+$ and let $w \in \mathcal{W}_i^+$ be contained in $x$. Then the event $\{\textrm{$x$ can be badly learned for budget $i$ due to infeasibility}\}$ is contained in $\skewm^S(\delta, w)$. Similarly, if $x \in \mathcal{Y}_i^+$ let $w \in \mathcal{Z}_i^-$ contain $x$; then the event $\{\textrm{$x$ can be badly learned for budget $i$ due to infeasibility}\}$ is contained in $\skewm^T(0, w)$. The reasoning for the event $\{\textrm{$x$ can be badly learned for budget $i$ due to value}\}$ is similar.
	
	The following is analogous to Lemma \ref{lemma:badWitness}.
	
	\begin{lemma} \label{lemma:badWitnessOTP}
		Suppose that, for all $i \in [m]$, there are witness sets for $(\mathcal{X}_i^+, \mathcal{X}_i^-, \mathcal{Y}_i^+, \mathcal{Y}_i^-)$ of size at most $M$. Then $\Pr(x^S \textrm{ is bad }) \le 8mM \exp\left(-\frac{\delta^2 s B}{136 n}\right)$.
	\end{lemma}


	\paragraph{Good witness sets.} We now construct witness sets of size at most $(O(\frac{K}{\delta} \log \frac{K}{\delta}))^m$, so Lemma \ref{lemma:goodSDOTP} will follow directly from Lemma \ref{lemma:badWitnessOTP}. The development mirrors that of Section \ref{sec:goodWitness}. Let $C_1, C_2, \ldots, C_K$ be a partition of the index set $[n]$ such that for all $j$, the columns $\{a^t\}_{t \in C_j}$ belong to the same 1-dimensional subspace.
	
	 Cover the interval $[0, B+m]$ with intervals $\{I_\ell\}_{\ell \in L}$, where $I_0 = [0, \frac{\delta B}{8 K})$ and $I_\ell = [\frac{\delta B}{8K} (1 + \frac{\delta}{8})^{\ell - 1}, \frac{\delta B}{8K} (1 + \frac{\delta}{8})^\ell)$ for $\ell > 0$ and $L = \{0, \ldots, \lceil \log_{1 + \delta/8} \frac{16K}{\delta} \rceil + 1\}$. Define $\mathcal{B}_{i,j}^\ell$ as the set of classifications $x \in \mathcal{X}|_{C_j}$ whose occupation $a_i(x)$ lies in the interval $I_\ell$. Finally, for $\ell \in L^K$, define the family of boxes $\mathcal{B}_i^\ell = \prod_j \mathcal{B}_{i,j}^{\ell_j}$.
	
	 Given $\ell \in L$, let $\underline{w}^\ell(j)$ be the smallest set in $\mathcal{X}|_{C_j}$ which has $a_i(\underline{w}^\ell(j)) \in I_\ell$ and for $\ell \in L^K$ define the set $\underline{w}^{\ell}$ as the union of the sets $\underline{w}^{\ell_j}(j)$'s (or equivalently, as the concatenation of the vectors $\underline{w}^{\ell_j}(j)$'s). Similarly, for $\ell \in L$ let $\overline{w}^\ell(j)$ be the largest set in $\mathcal{X}|_{C_j}$ which has $a_i(\overline{w}^\ell(j)) \in I_\ell$ and for $\ell \in L^K$ define the set $\overline{w}^\ell$ as the union of the sets $\overline{w}^{\ell_j}(j)$'s.
	
	 Now we construct the witness sets as before. Set $\mathcal{W}_i^+ = \{\underline{w}^\ell : a_i(\underline{w}^\ell) \ge (1 - \delta)B + \frac{\delta B}{4}, \mathcal{B}_i^\ell \cap \mathcal{X} \neq \emptyset\}$, set $\mathcal{Z}_i^+ = \{\underline{w}^\ell : a_i(\underline{w}^\ell) \ge (1 - 2\delta)B - \frac{3\delta B}{4}, \mathcal{B}_i^\ell \cap \mathcal{X} \neq \emptyset\}$, set $\mathcal{W}_i^- = \{\overline{w}^\ell : a_i(\overline{w}^\ell) \le (1 - 2\delta)B - \frac{\delta B}{4}, \mathcal{B}_i^\ell \cap \mathcal{X} \neq \emptyset\}$ and finally set $\mathcal{Z}_i^- = \{\overline{w}^\ell : a_i(\overline{w}^\ell) \le (1 - \delta)B + \frac{3\delta B}{4}, \mathcal{B}_i^\ell \cap \mathcal{X} \neq \emptyset\}$.
	
	 Following the same steps as in the proof of Lemma \ref{lemma:wWitness}, one can check that $(\mathcal{W}_i^+, \mathcal{W}_i^-, \mathcal{Z}_i^+, \mathcal{Z}_i^-)$ are \emph{witness sets} for $(\mathcal{X}_i^+, \mathcal{X}_i^-, \mathcal{Y}_i^+, \mathcal{Y}_i^-)$. Moreover, the proof of Lemma \ref{lemma:sizeP} can be used to show that, for a fixed $i \in [m]$, at most $(e \frac{K}{\delta} \log \frac{K}{\delta} )^m$ of the $\mathcal{B}_i^\ell$'s contain an element of $\mathcal{X}$, which then imposes the same upper bound on the size of the witness sets. This concludes the proof of Lemma \ref{lemma:goodSDOTP}.

	\medskip
	\begin{proofof}{Lemma \ref{lemma:sDeltaOtp}} Let $x$ be the solution returned by $(s, \delta)$-\otp and let $\mathcal{E}$ denote the event that $x^S$ is good. For any scenario in $\mathcal{E}$, we have $x_{\sigma(t)} = x^S_{\sigma(t)}$ for all $t = s+1, s+2, \ldots, 2s$. Therefore, we get that
		\begin{align}
			\E\left[\sum_{t = 1}^{2s} \pi_{\sigma(t)} x_{\sigma(t)}\right] &\ge 		   \E\left[\sum_{t = 1}^{2s} \pi_{\sigma(t)} x_{\sigma(t)} \mid \mathcal{E} \right] \Pr(\mathcal{E}) \notag \\
			& \ge \E\left[\sum_{t = 1}^{2s} \pi_{\sigma(t)} x_{\sigma(t)}^S \mid \mathcal{E}\right] \Pr(\mathcal{E}) - \E[\OPT(s) \mid \mathcal{E}] \Pr(\mathcal{E}) \notag \\
	&\ge \E\left[\sum_{t = 1}^{2s} \pi_{\sigma(t)} x_{\sigma(t)}^S \mid \mathcal{E}\right] \Pr(\mathcal{E}) - \E[\OPT(s)]. \label{eq:sDeltaOpt1}
		\end{align}
		
		To lower bound the first term in the right hand side we use again the definition of $\mathcal{E}$:
		\begin{align*}
			 \E\left[\sum_{t = 1}^{2s} \pi_{\sigma(t)} x_{\sigma(t)}^S \mid \mathcal{E} \right] \ge (1 - 3 \delta) \E[\OPT(2s) \mid \mathcal{E}] \Pr(\mathcal{E})
		\end{align*}
		and
		\begin{align*}
			 \E[\OPT(2s)] = \E[\OPT(2s) \mid \mathcal{E}] \Pr(\mathcal{E}) + \E[\OPT(2s) \mid \overline{\mathcal{E}}] \Pr(\overline{\mathcal{E}}) \le \E[\OPT(2s) \mid \mathcal{E}] \Pr(\mathcal{E}) + \delta^2 \OPT,
		\end{align*}		
		where the last inequality uses Lemma \ref{lemma:goodSDOTP}. Combining the previous two inequalities give that $\E\left[\sum_{t = 1}^{2s} \pi_{\sigma(t)} x_{\sigma(t)}^S \mid \mathcal{E}\right] \ge (1 - 3 \delta) \E[\OPT(2s)] - \delta^2 \OPT$, and the result follows from equation \eqref{eq:sDeltaOpt1}.
	\end{proofof}


	\section{Proof of Theorem \ref{thm:expValueDPA}}
	
	Let LP1 denote the LP with columns $(\pi_t, \tilde{a}^t)$ and right-hand side $\tilde{B} = (1-\epsilon)B$ and LP2 denote the LP with columns $(\pi_t, a^t)$ and right-hand side $B$. We show that Robust \dpa returns a $(1 - 21.5\epsilon)$-approximation for LP1, and the theorem will follow from Lemma \ref{lemma:robust}.
	
	First we show that the returned solution $x$ is feasible for LP1. By definition of the algorithm, $a_j(x^i) \le \epsilon 2^i \tilde{B}$ for all $i,j$. By linearity, $a_j(x) = \sum_i a_j(x^i) \le \epsilon \tilde{B} \sum_{i = 0}^{\log(1/\epsilon) - 1} 2^i \le \tilde{B}$.
	
	In order to verify the value of the returned solution, we first show that $\frac{\delta^2 s B}{n} \ge \Omega(m \ln \frac{K}{\delta})$ in every call to $(s,\delta)$-\otp made by Robust \dpa. As in Section \ref{sec:robustOTP}, the columns $\tilde{a}^t$'s belong to at most $K = O(\frac{m}{\epsilon})^m$ 1-dim subspaces. Since $B \ge \Omega(\frac{m^2}{\epsilon^2} \ln \frac{m}{\epsilon})$, we have that for each $i = 0, \ldots, \log(1/\epsilon)-1$ setting $s = \epsilon 2^i n$ and $\delta = \sqrt{\epsilon/2^i}$ satisfies the expression $\frac{\delta^2 s B}{n} \ge \Omega(m \ln \frac{K}{\delta})$.
	
	Then applying Lemma \ref{lemma:sDeltaOtp} we get that for all $i = 0, \ldots, \log(1/\epsilon)-1$, $\E[\pi x^i] \ge (1 - 3 \sqrt{\frac{\epsilon}{2^i}}) \E[\OPT(\epsilon 2^{i+1} n)] - \E[\OPT(\epsilon 2^i n)] - \frac{\epsilon \OPT}{2^i}$. By linearity of the objective value and of expectations
	\begin{align*}
		\E[\pi x] = \sum_i \E[\pi x^i] \ge - \E[\OPT(\epsilon n)] - \sum_{i = 0}^{\log(1/\epsilon) - 2} \left(3 \sqrt{\frac{\epsilon}{2^i}}\right) \E[\OPT(\epsilon n 2^{i + 1})] + (1 - 3 \sqrt{2} \epsilon - \epsilon) \OPT.
	\end{align*}
	
	Lemma 2.4 of~\cite{agrawal} states that $\E[\OPT(s)] \le \frac{s}{n} \OPT$ for all $s \ge 0$. Employing this observation, we get
	\begin{align*}
		\E[\pi x] \ge \OPT - \epsilon \OPT \left[3 \sqrt{2} + 2 + 3 \sqrt{\epsilon} \sum_{i = 0}^{\log(1/\epsilon) - 2} 2^{i/2 + 1} \right].
	\end{align*}
	Since the summation in the expression can be upper bounded by $\frac{2 \sqrt{2}^{\log(1/\epsilon)}}{\sqrt{2} - 1} \le \frac{5}{\sqrt{\epsilon}}$, we get that $\E[\tilde{\pi} x] \ge (1 - 21.5 \epsilon) \OPT$. This concludes the proof of the theorem.

\end{document}